\documentclass[aps,pra,superscriptaddress,showpacs,twocolumn,10pt]{revtex4-1}
\usepackage[colorlinks=true,linkcolor=blue,urlcolor=blue,citecolor=blue]{hyperref}
\usepackage{color}
\usepackage{amssymb}
\usepackage{graphicx}
\usepackage{amsmath}
\usepackage{amsthm}
\usepackage{bbm}
\usepackage{braket}

\begin{document}

%\preprint{APS/123-QED}

\title{Charge Transfer in Ultracold Rydberg-Ground State Atomic Collisions}% Force line breaks with \\

\author{Samuel Markson}
 \email{smarkson@cfa.harvard.edu}
\affiliation{%
 ITAMP,  Harvard-Smithsonian Center for Astrophysics 60 Garden St., Cambridge, MA 02138, USA}
\affiliation{Physics Department, University of Connecticut, 2152 Hillside Rd., Storrs, CT 06269-3046, USA
}
\author{H. R. Sadeghpour}%
 \email{hrs@cfa.harvard.edu}
\affiliation{%
 ITAMP,  Harvard-Smithsonian Center for Astrophysics 60 Garden St., Cambridge, MA 02138, USA}
%\affiliation{
%Physics Department\\
%University of Connecticut\\
%2152 Hillside Rd.\\
%Storrs, CT 06269-3046, USA
%}{UCONN

\date{\today}% It is always \today, today,

\begin{abstract}
In excited molecules, the interaction between the covalent Rydberg and ion-pair channels forms a unique class of excited Rydberg states, in which the infinite manifold of vibrational levels are the equivalent of atomic Rydberg states with a heavy electron mass. 
Production of the ion pair states usually requires excitation through one or several interacting Rydberg states; these interacting channels are pathways for loss of flux, diminishing the rate of ion pair production.
Here, we develop an analytical, asymptotic charge transfer model for the interaction between ultracold Rydberg molecular states, and employ this method to demonstrate the utility of off-resonant field control over the ion pair formation, with near unity efficiency. 
%In this work, we explore control of non-adiabatic transitions to creating such exotic species. We study time-independent techniques for increasing efficiencyof forming heavy Rydberg states with large dipole moments.
\end{abstract}
\maketitle

\section{Introduction}
Ultracold atomic systems allow for precise control via laser and static fields and are employed for simulating strongly-correlated many-body condensed matter and optical systems, for performing chemistry in the quantum regime, and for executing quantum computing protocols ~\cite{bloch_many-body_2008,regal_creation_2003,carr_cold_2009}. 
Excitations in ultracold atomic traps have ushered in the Rydberg blockade regime~\cite{lukin_dipole_2001,saffman_quantum_2010} (with promise for quantum information processing and quantum bit operations), ultracold plasmas~\cite{castro_role_2009} (with application in recombination, ion crystal order and heating), and ultra long range Rydberg molecules~\cite{greene_creation_2000,bendkowsky_observation_2009} (for studies of few-body molecular systems, symmetry breaking~\cite{li_homonuclear_2011,,booth_production_2015} and coherent control~\cite{rittenhouse_ultracold_2010}).

Another class of molecular states, the ion pair states, form channels when the covalent Rydberg channels couple to the long-range ion-pair potential. These atomic cation-anion pair states share several properties with ionic molecules; an infinite spectrum of vibrational levels which follow a Rydberg progression with a heavy electron mass, and large permanent electric dipole moments. They are also long-range states and have typically negligible Franck-Condon (FC) overlap with the usual short-range molecular levels. These heavy Rydberg states (HRS) have been experimentally observed in several molecular species, relying on excitation from bound molecular levels ~\cite{vieitez_observation_2008,vieitez_spectroscopic_2009,mollet_dissociation_2010,ekey_spectroscopic_2011}.

The prevailing issue with indirect excitation of bound molecules, such as in H$_2$ and Cl$_2$~\cite{vieitez_observation_2008,vieitez_spectroscopic_2009,mollet_dissociation_2010,ekey_spectroscopic_2011}, is that it is not {\it a priori} possible to identify a set of long-lived intermediate heavy Rydberg states to which ion the pair states couple. 
In a recent work~\cite{kirrander_approach_2013}, it was proposed to directly pump long-lived HRS from ultracold Feshbach molecular resonances, just below the avoided crossings between the covalent potential energy curves and the ion-pair channel. The predominant excitation to HRS occurs near the avoided crossings, because the non-adiabatic mixing allows for favorable electronic transitions to the HRS.
When the nuclear HRS wave function peaks at the classical turning point, the internuclear FC overlap increases.

In this work, we develop an analytical, but asymptotic model for one-electron transfer, merging single-center potentials for the Rb atom, and demonstrate efficient field control over the rate of ion pair formation. 
Adiabatic potential energy curves are calculated along with the radial non-adiabatic coupling and dipole transition matrix elements. We compare our adiabatic potential energies with the Born-Oppenheimer (BO) potential energy curves from Ref.~\cite{park_theoretical_2001}. 
We find that with modest off-resonant external fields, we can alter the avoided crossing beween the covalent HRS and ion pair channels, and modify the behavior of the nuclear wavefunction at the classical turning points; hence, control the FC overlaps and rate of ion pair formation.

%We directly excite to the Rb(5s)+Rb(7p) and $\text{Rb}^{+}({}^1\text{S})+\text{Rb}^{-}({}^1\text{S})$ channels. 
This excitation occurs at larger internuclear separations, where the overlap and transition dipole moments between the ground and ion pair states are significant.  
The method proposed in~\cite{kirrander_approach_2013} requires adiabatic rapid passage or multiphoton transitions to enhance this excitation, while in the present work, only off-resonant fields must be employed to increase the efficiency of ion pair production.  

\begin{figure}
  \includegraphics[width=0.45\textwidth]{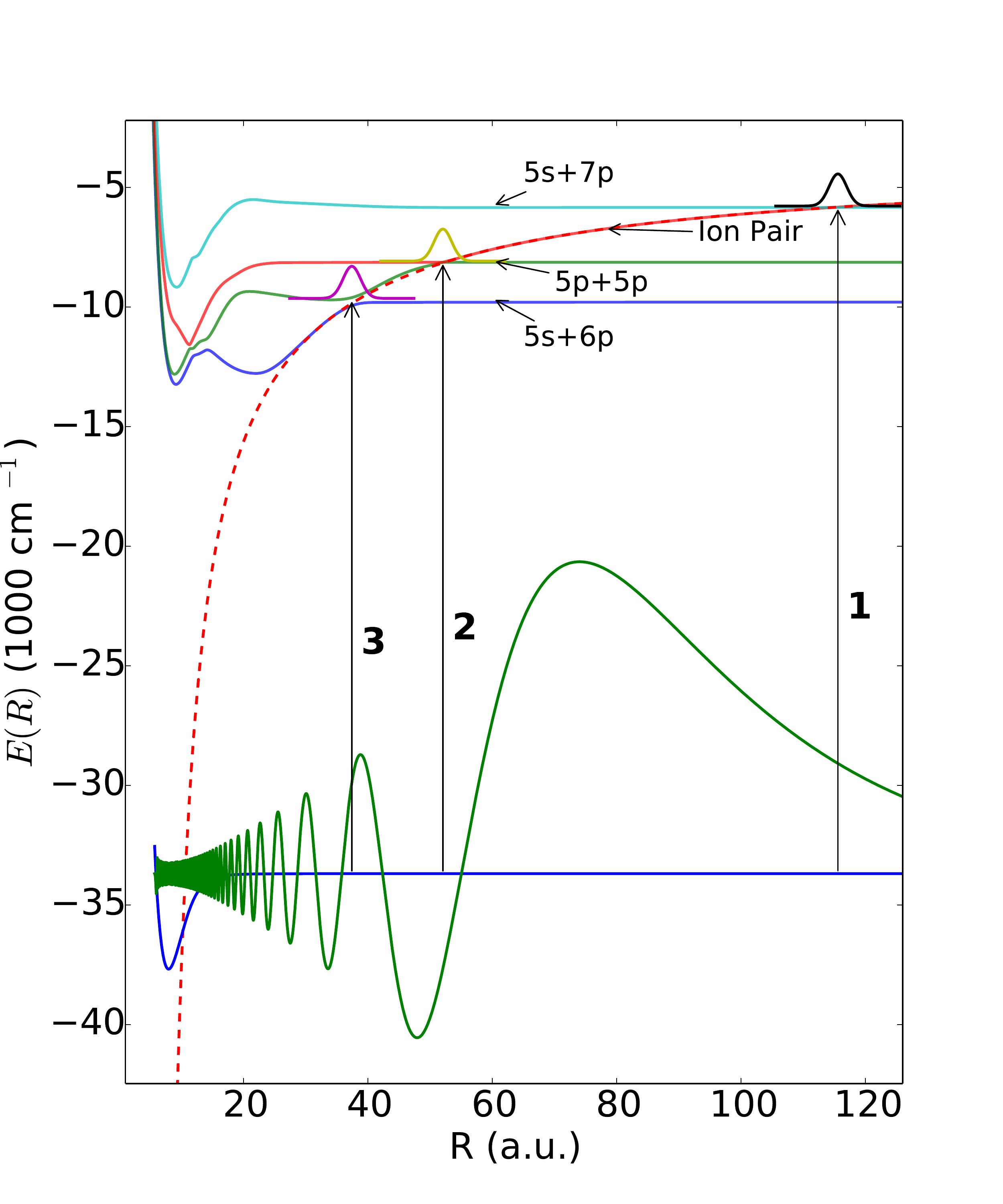}
  \caption{Scheme from \cite{kirrander_approach_2013} relied on excitation to the Rb(5s)+Rb(6p)/Rb$^+$+Rb$^-$ (ion pair) or Rb(5p)+Rb(5p)/ion pair crossings (excitation 2 and 3).  Here, we focus on the Rb(5s)+Rb(7p)/ion pair crossing (excitation 1). Subsequent to excitation, both schemes from \cite{kirrander_approach_2013} require multiphoton excitation or chirped pulse rapid adiabatic passagee.  Excitation 1, directly to the Rb(5s)+Rb(7p)/ion pair crossing, can more easily be controlled with dc electric fields.  Additionally, the Franck-Condon factors for this excitation are favorable, as demonstrated by the $\nu=124$ Feshbach molecule vibrational wavefunction. The dashed curve is the ion pair potential.}
\end{figure} 

\section{\label{sec:level2}Methodology and Results}
The present charge transfer model relies on one active electron participating in the process of ion pair excitation. For the charge transfer to occur, the Rb valence electron is ionized from one center, separated from the other neutral center by a distance $R$. The ionization process is calculated in the one-electron effective potential model of Ref.~\cite{marinescu_dispersion_1994}. The interaction
 of the electron with the neutral Rb atom is modeled by the short-range scattering of the low-energy electron from the ground-state Rb atom. This is done with a symmetric inverse hyperbolic cosine (Eckart) potential.

The two-center Hamiltonian matrix elements are constructed by expanding in a basis of atomic Rb Rydberg orbitals and the single bound orbital in the Eckart potential. The resulting generalized eigenvalue equation is solved for the adiabatic potential energy curves, radial coupling and electronic dipole matrix elements. Our control scheme hinges on modifying the avoided crossing gap between the 5s+7p and ion pair state by applying an off-resonant field, thereby tuning the Landau-Zener crossing probabilities between the ion pair and covalent states.
%While the mechanism here is hypothesized to be that behind both ``bond-hardening'' and ``bond-softening'' ~\cite{townsend_stark_2011}, the effect in the Rydberg case is enhanced.  To estimate the response of an avoided crossing to an applied field we devise a pseudopotential that should accurately and inexpensively predict the geometry of the avoided crossing and be capable of generating relevant quantities (dipoles, non-adiabatic coupling matrix elements) in a neighborhood of those crossings.  

%We assume a single active electron, and therefore assume that the other, ground state electron remains frozen to its rubidium center, and may therefore be absorbed into the pseudopotential itself.  
The two-center potential for the Rb$_2$ Rydberg-excited molecule is the sum of two single-center potentials,
\begin{equation}
V_{\text{model}} (r_a, r_b|R) = V_1 (r_a|R) + V_2 (r_b|R) 
\label{eq:potential}
\end{equation}
where $V_1$ represents the potential felt by the valence electron due to the Rb core, and the pseudopotential $V_2$ that of the binding to the neutral rubidium atom; $r_a$ and $r_b$ represent the distances between the electron and the cation/neutral atom, respectively.  

We require that the potential in Eq.~\ref{eq:potential} correctly reproduce the asymptotic dissociation energies for all the relevant Rydberg and ion-pair states \cite{NIST_table}.  
We further enforce that the known energy-independent $e^{-}\text{-Rb}$ scattering length and the Rb electron affinity are reproduced, i.e. $a_{sc} =-16.1$ a.u.~\cite{bahrim_3se_2001} and EA=-0.01786 a.u. \cite{NIST_table}.

The valence-electron potential, $V_1$~\cite{marinescu_dispersion_1994} correctly reproduces the observed atomic energies ~\cite{NIST_table},
\begin{equation}
V_1(r_a)= - \frac{Z_l (r_a)}{r_a} - \frac{\alpha_c}{2 r_a^4} (1- e^{-(r_a/r_c)^6}) 
\label{eq:leftpotential}
\end{equation}
where
\[Z_l(r_a)=1+(z-1) e^{-\alpha_1 r_a} -r (a_3 + a_4 r_a) e^{-a_2 r_a} \]
with $z$ being the nuclear charge.  The terms of the potential account for screening of the nuclear charge due to core electrons and effect of core polarizability. 

For the neutral rubidium electron affinity, we use the radial Eckart potential~\cite{eckart_penetration_1930}.
\begin{equation}
V_2(r_b)=V_0 \cosh^{-2} {(\frac{r_b}{r_0})}
\label{eq:rightpotential}
\end{equation}
where $V_0$ and $r_0$ are parameters related in a set of coupled equations to the electron affinity and $e^{-}\text{-Rb}$ scattering length; see appendix A.  

The Hamiltonian matrix is constructed from a suitable basis for each center and may be conveniently represented in prolate spheroidal coordinates  
\[r=\frac{R}{2} (\xi + \kappa \eta),\ \cos{\theta} = \frac{1+\kappa \xi \eta}{\xi + \kappa \eta} \]
where $\kappa = +1(-1)$ represents the radial coordinate of the left(right) centers, $\xi \in [1,\infty),\ \eta \in [-1,1]$.

We solve the generalized eigenvalue problem:
\begin{equation}
\mathbf{H} \vec{\Psi} = E(R) \mathbf{S} \vec{\Psi}\label{eq:scheq}
\end{equation}
where $\mathbf{H}$ and $\mathbf{S}$ are the Hamiltonian and overlap matrices in the truncated basis:
\begin{equation*}
\mathbf{H}_{jj'}=\bra{\phi_j^{(a)}} H_0^{(a)}+V_2 \ket{\phi_j^{(a)}}
\tag{4a}
\end{equation*}
\begin{equation*}
\mathbf{H}_{kk'}=\bra{\phi_k^{(b)}} H_0^{(b)}+V_1 \ket{\phi_k^{(b)}} 
\tag{4b}
\end{equation*}
\begin{equation*}
\mathbf{H}_{jk}=\bra{\phi_j^{(a)}} H_0^{(b)}+V_1 \ket{\phi_k^{(b)}} = \bra{\phi_k^{(b)}} H_0^{(a)} + V_2 \ket{\phi_j^{(a)}}
\tag{4c}
\end{equation*}
\begin{equation*}
\mathbf{S}_{jj'}=\braket{\phi_j^{(a)} | \phi_j'^{(a)}} = \delta_{j j'}
\tag{4d}
\end{equation*}
\begin{equation}
\mathbf{S}_{kk'}=\braket{\phi_k^{(b)} | \phi_k'^{(b)}} = \delta_{k k'}
\tag{4e}
\end{equation}
\begin{equation*}
\mathbf{S}_{jk}=\braket{\phi_j^{(a)} | \phi_k^{(b)}} 
\tag{4f}
\end{equation*}

The full wave function is comprised of $\{\phi_j^{(a)}(\frac{R}{2} (\xi + \eta)\}$ centered on $V_1$, and $\{\phi_k^{(b)}(\frac{R}{2} (\xi - \eta)\}$ centered on $V_2$, i. e. $\Psi_i= \sum_j^{n_a} c_{ij} \phi_j^{(a)} + \sum_k^{n_b} c_{i k} \phi_k^{(b)}$.
 %$\{\phi_j^{(a)}(\frac{R}{2} (\xi + \eta)\}$ are the basis vectors radially symmeric about the first center, with $\{\phi_k^{(b)}(\frac{R}{2} (\xi - \eta)\}$ those about the second.
The truncated basis set contains atomic orbitals, $\{\phi_j^{(a)}\} = \{(5\text{-}9)s,(5\text{-}11)p,(4\text{-}6)d\}$, and the short-range wave function for scattering the of electrons from $V_2$, 
%that is, the above atomic orbitals, as calculated using the pseudopotentials from ~\cite{marinescu_dispersion_1994}, calculated numerically via the Numerov-Cooley algorithm, and
$\{\phi_k^{(b)}\} = \{\cosh^{-2 \lambda}\frac{r_b}{r_0} \sqrt{z}\ {}_2 F_1(0,-2 \lambda,\frac{3}{2},z)\}$, with 
$z=-\sinh^2{(\frac{r_b}{r_0})},\ \mbox{and}\ \lambda=0.0321816$; see Appendix~\ref{sec:eckappendix}. ${}_2 F_1(a,b;c;z)$ is the hypergeometric function ~\cite{bateman_manuscript_project_tables_1954}.

\begin{figure}
 \centering
  \includegraphics[width=0.5\textwidth]{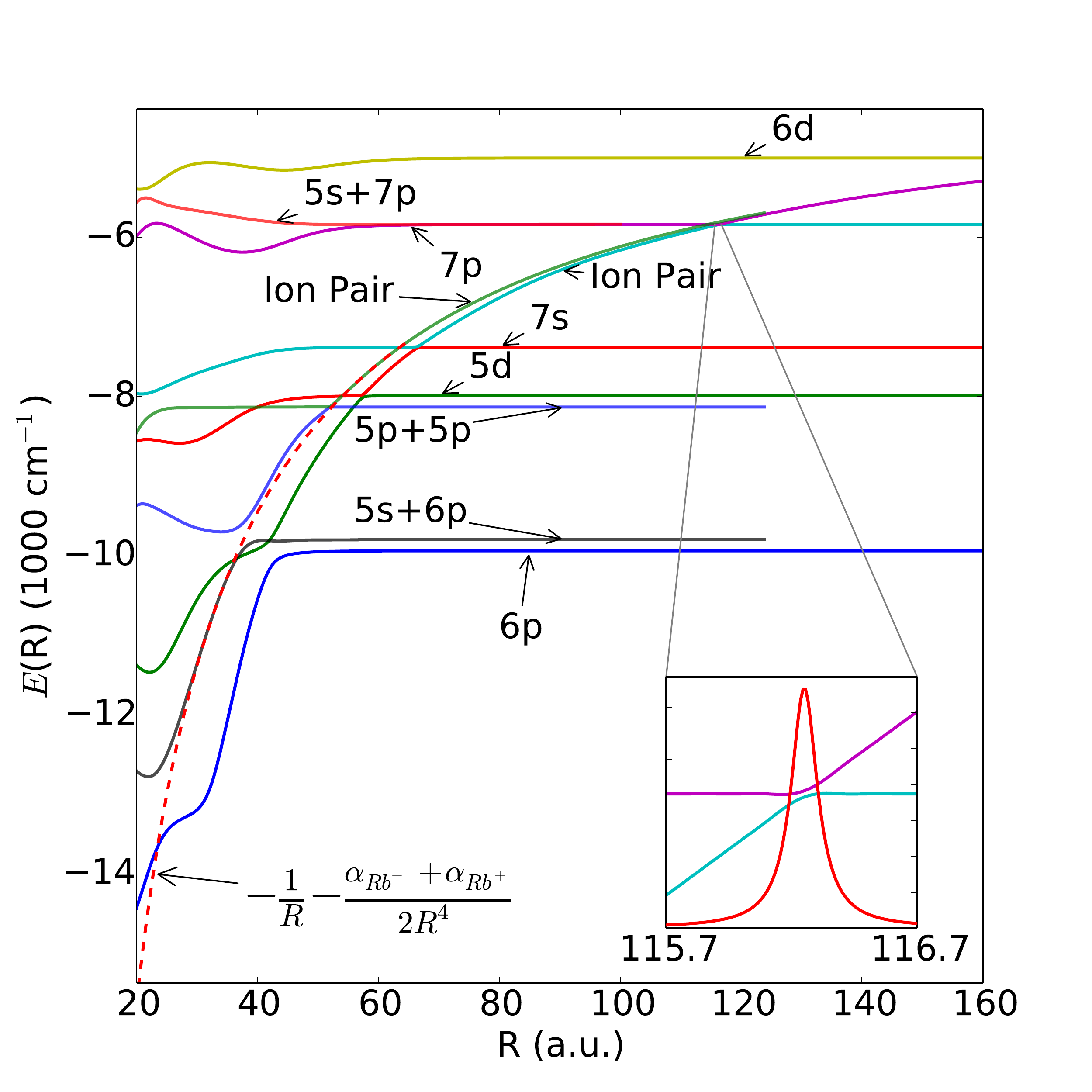}
 \caption{Nuclear potential energy curves, overlaid with calculations from other publications.  Calculated curves are well-behaved in the large R limit.  Inset shows avoided crossing at $R_c=116.2$, along with non-adiabatic coupling matrix element between Rb(5s)+Rb(7p)/ion pair channels.  Rb(5s)+Rb(6p), Rb(5p)+Rb(5p), and ion pair curves taken from Rb(5s)+Rb(6p), Rb(5p)+Rb(5p), Rb(5s)+Rb(7p), and ion pair curves taken from ~\cite{park_theoretical_2001} ~\cite{bellos_upper_2013}.  Ion pair curve is fitted to $-\frac{1}{R}-\frac{\alpha_{\text{Rb}_{-}}+\alpha_{\text{Rb}_{+}}}{2 R^4}$, where $\alpha_{\text{Rb}_{-}}=526.0$ a.u ~\cite{fabrikant_polarizability}, $\alpha_{\text{Rb}_{+}}=9.11$ a.u  ~\cite{clark_polarizability} are the polarizabilities of the anion and cation respectively. }
  \label{fig:potential}
 \end{figure}

One major feature of this asymptotic approach is that the permanent and transition dipole, and non-adiabatic radial coupling matrix elements can now be calculated from the eigenstates of Eq.~\ref{eq:scheq}. Full details are available in Appendix~\ref{sec:eckappendix}.  The adiabatic potential energy curves, $E(R)$, are shown in Fig.~\ref{fig:potential}. By construction, these adiabatic potentials correlate to the  asymptotic dissociation energies for the Rydberg and ion pair states and have avoided crossings between covalent and ion pair channels. The Rb$_2$ BO potential energies in the region of Rb(5s)+Rb(6p) dissociation energy are superposed on the adiabatic potentials for comparison. The non-adiabatic coupling matrix element between the ion pair channel with the molecular Rydberg curve dissociating to Rb(5s)+Rb(7s) is shown in the inset of Fig.~\ref{fig:potential}.

\section{Field Control Covalent/Ion Pair Channels}
The avoided crossing in the interaction of ion pair channel (Rb$^+$+Rb$^-$) and Rydberg channel (Rb(5s)+Rb(7p)) is nearly diabatic, i. e. the covalently populated vibrational states will predominantly dissociate to neutral atoms with little possibility of forming HRS and ion pair states. This is reflected in the narrowness of the avoided crossing and the strength of the radial non-adiabatic matrix element. The two channels have non-zero dipole transition moments, so in an external field, they will mix and modify the transition probability for populating ion pair states. 

\begin{figure}
 \centering
  \includegraphics[width=0.5\textwidth]{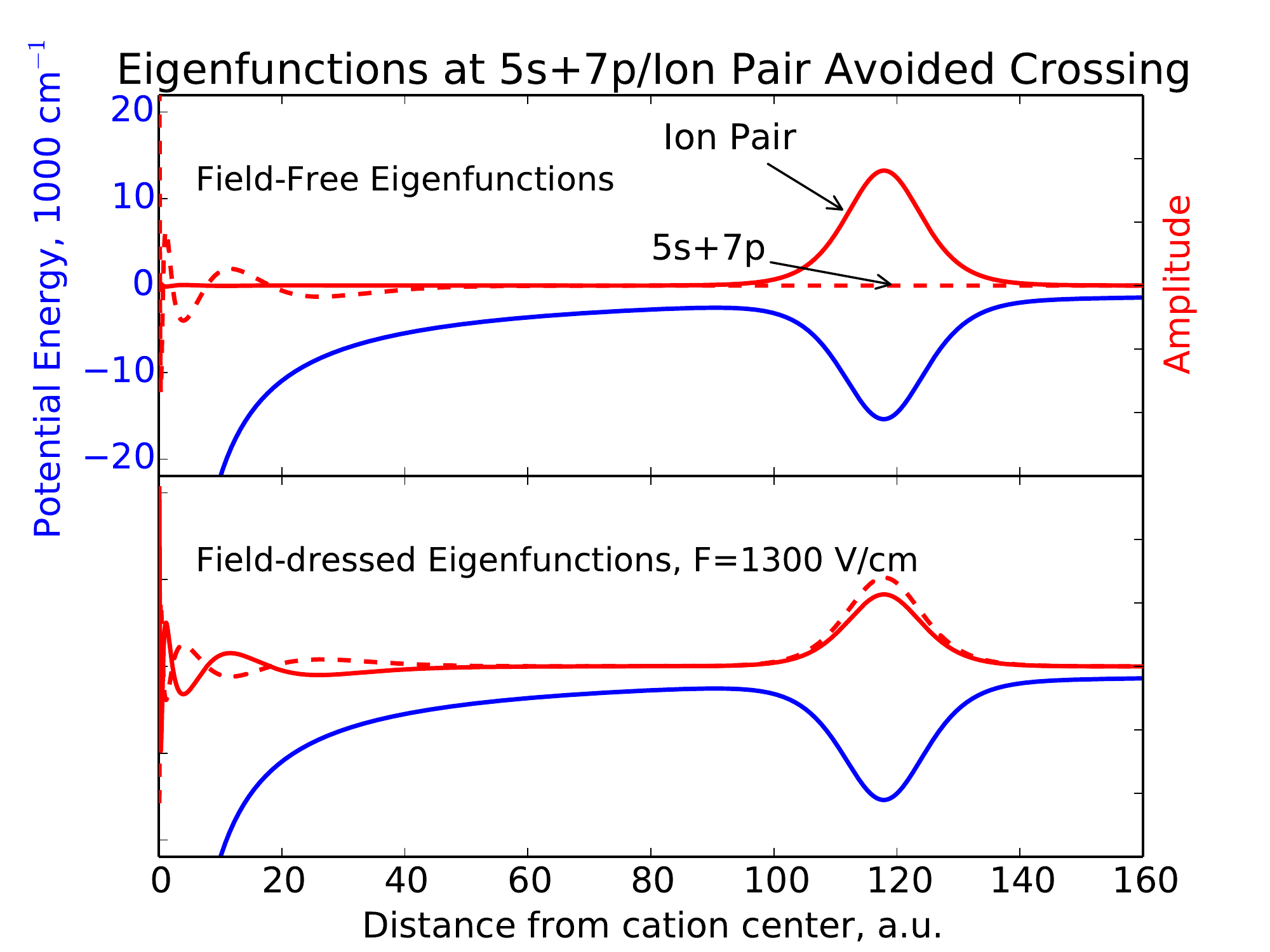}
  \caption{Hybridization of the wavefunctions at the avoided crossing by the field.  The upper panel show the cross section of the field-free electronic wavefunction along the internuclear axis---states associated with the ion pair and Rb(5s)+Rb(7p) channels have very little overlap.  The lower panel is field-dressed ($\mathbf{F}=1300$ V/cm, showing significant overlap of the wavefunctions associated with the two channels.  The blue line is the pseudopotential along the internuclear axis for $R=116.2$ a.u.}
 \label{fig:hybrid}
 \end{figure}

Near the avoided crossing, the electronic wave function becomes hybridized in the field, as in the first order of perturbation theory,
$\ket{\psi} = \ket{\psi_0} + \sum_{k \neq 0} \frac{\bra{ \psi_k}  - \mathbf{F} \cdot \mathbf{z} \ket{\psi_0}}{E_0-E_k} \ket{\psi_k}$, where $\ket{\psi_k}$ are the dipole-allowed states which couple to the initial state, $\ket{\psi_0}$. This hybridization is demonstrated in Fig.~\ref{fig:hybrid} for the two channels of interest. When the field is off, the wave function amplitude in the covalent Rydberg channel peaks near the avoided crossing. With the field on, the two amplitudes become comparable.

The probability that ion-pair states survive the single-pass traversal through the avoided crossing region in Fig. ~\ref{fig:potential} is given approximately by the Landau-Zener-St\"{u}ckelberg formula:
\begin{equation}
P_{\text{ad}}=1-\exp{[-2 \pi \frac{a^2}{|v \frac{\partial (E_2-E_1)}{\partial R} |}]} 
\label{eq:lzform}
\end{equation}
where $v = \sqrt{\frac{2(E- V(R_c))}{m}}$ is the velocity of the wavepacket at crossing, $R=R_c$, $a$ is the off-diagonal element coupling the two states at $R_c$ (half the crossing size in the fully diagonalized picture), and $\frac{\partial (E_2(R)-E_1(R))}{\partial R}|_{R=R_c}$ is the relative slope of the intersecting curves at $R=R_c$.  

 \begin{figure}[h!]
 \centering
  \includegraphics[width=0.5\textwidth]{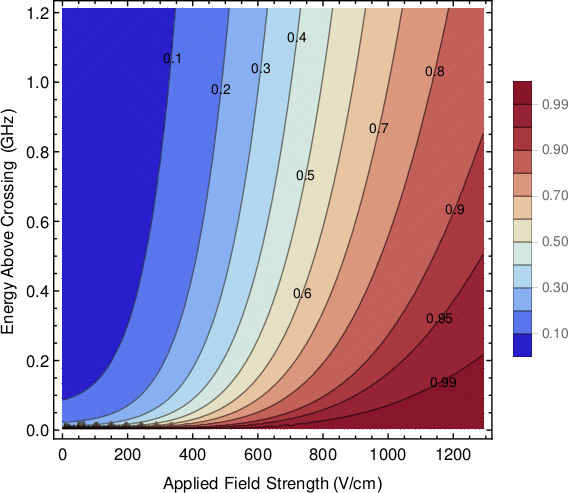}
  \caption{How the ion-pair survival probability may be controlled with modest off-resonant electric fields:  Perturbation theory becomes unreliable for \textbf{F} $>1700$ V/cm, and classical ionization occurs for \textbf{F} $>5$ MV/cm.  The vertical width of contour regions gives a measure of the survival probability above a desired threshold for a laser-pulse-excited vibrational wavepacket.  For example, at $\mathbf{F} \simeq 1000 $V/cm., more than 2 $\sigma_{\text{std}}$ will survive if the laser linewidth can be constrained to $\sim150$ MHz. }
  \label{fig:lzc}
 \end{figure}

Fig. ~\ref{fig:lzc} illustrates how the ion=pair probability can be controlled with modest electric fields.  
Each curve represents a contour of cross-sectional survival probability for a vibrational wavepacket excited to a certain energy above the crossing threshold at a given field strength.  For example, at field strengths of $1000 $V/cm., more than 2 $\sigma_{\text{std}}=95.45\%$ will survive if the laser linewidth can be constrained to $\sim150$ MHz.

As the loss probability of a given vibrational level goes at weak field strength as $e^{-\mathbf{F}^4}$, even relatively weak fields have a profound effect on ion pair state survival.  

\section{Conclusion}
We have shown that application of weak off-resonant electric fields provide a simple yet potent means of drastically increasing the survivability of the ion pair states in Rb dimers.  To this end, we have introduced a simple and intuitive analytical pseudopotential which reproduces the correct asymptotic behavior of the dimer.  
This model provides an easy and computationally straightforward way for us to calculate dipole and non-adiabatic coupling matrix elements for excited states of alkali dimers.

This work was supported by the National Science Foundation through a grant for the Institute for Theoretical Atomic, Molecular, and Optical Physics at Harvard University and Smithsonian Astrophysical Observatory; S. Markson was additionally supported through a graduate research fellowship through the National Science Foundation.

\bibliography{ionbib}
\bibliographystyle{apsrev4-1-etal}
%\printbibliography

\appendix
\section{The Eckart potential}
\label{sec:eckappendix}
\subsection{Bound states}
%The following is taken largely from Gol'dman, Krivchenkov \cite{gol?dman_problems_2006}.
Taking the Schr\"{o}dinger equation with $V(r)$ from ~\ref{eq:rightpotential}, we make the substitutions~\cite{gol?dman_problems_2006}
\[\psi = (\cosh \frac{r}{r_0})^{-2 \lambda} u,\ \text{where}\ \lambda=\frac{1}{4} (\sqrt{\frac{8 \mu V_0 {r_0}^2}{ \hbar^{2}} +1} -1) \]
The Schr\"{o}dinger equation then becomes
\begin{equation}
\frac{d^2 u}{dr^2} - \frac{4 \lambda}{r_0} \tanh{(\frac{r}{r_0})} \frac{du}{dr} + \frac{4}{{r_0}^2} (\lambda^2-\chi^2) u = 0 
\label{eq:transformedeckarteq}
\end{equation}
with
\[\chi = \sqrt{-\frac{\mu E {r_0}^2}{2 \hbar^2}}\]
Letting $z=-\sinh^2 (\frac{r}{r_0})$ leads us to the hypergeometric equation
\begin{equation}
z(1-z) \frac{d^2 u}{dz^2} + (\gamma-(\alpha + \beta +1) z ) \frac{du}{dz} - \alpha \beta u = 0
\label{eq:hypergeo}
\end{equation}
where $\gamma=\frac{1}{2},\ \alpha= \chi - \lambda,\ \beta=-\chi -\lambda$.

In spherical coordinates, only odd hypergeometric solutions are valid:
\begin{equation}
u=\sqrt{z}\ F (-\lambda + \chi + \frac{1}{2},-\lambda -\chi+\frac{1}{2};\frac{3}{2};z) 
\label{eq:hgl}
\end{equation}
By enforcing asymptotic boundary conditions for the bound states ($u\rightarrow 0$ as $ r \rightarrow \infty$), we have that
\[\lambda - \chi = n+\frac{1}{2},\ n=0,1,2,\cdots \]
giving the set of bound states 
\begin{equation}
\psi_n=(\cosh \frac{r}{r_0})^{-2 \lambda} u_n 
\label{eq:boundstates}
\end{equation}
where
\[u_n=N \sinh( \frac{r}{r_0}) F(-n,-2\lambda +n+1;\frac{3}{2};-\sinh^2( \frac{r}{r_0}))\]
(N being a normalizing constant), with corresponding energies 
\begin{equation}
E_n = -\frac{2 \hbar^2}{\mu {r_0}^2} (\frac{1}{4} \sqrt{\frac{8 \mu V_0 {r_0}^2}{ \hbar^{2}} +1} -n -\frac{3}{4})^2
\label{eq:eckarteigenenergies}
\end{equation}

\subsection{Scattering States}
With $E=\frac{\hbar^2 k^2}{2 \mu}$, $\chi^2 = -\frac{\mu E {r_0}^2}{2 \hbar^2} \rightarrow i k r_0 = 2 \chi$ \cite{gol?dman_problems_2006}, and

\begin{multline*}
 u(r) = N \sinh{( \frac{r}{r_0})} F(-\lambda + \frac{i k r_0}{2}+\frac{1}{2}, \\ -\lambda-\frac{i k r_0}{2}+\frac{1}{2}, \frac{3}{2},-\sinh^2 \frac{r}{r_0})
\end{multline*}
fulfilling the boundary conditions $$\lim_{r\rightarrow 0}u(r)=0,\ \lim_{r\rightarrow \infty} u(r) \simeq \sin ( k r + \delta_0) $$
where $\delta_0$ is the usual phase shift.  

We note that $F(\alpha,\beta; \gamma; 0)=1,\ \forall \alpha,\beta,\gamma$, and that $$\lim_{r \to \infty} { \sinh^2} (\frac{r}{r_0}) \approx { \frac{1}{2}} e^{\frac{2 r}{r_0}} $$

Using the asymptotic behavior of the Gaussian hypergeometric functions (see ~\cite{bateman_manuscript_project_tables_1954}, p. 108, equation 2).
\[u(r) = A e^{i k r} + B e^{-i k r} \]
where
\[ A= 2^{-i k r_0} \frac{\Gamma(i k r_0)}{\Gamma(\frac{1-2 \lambda+i k r_0}{2}) \Gamma(1+\frac{2 \lambda+ i k r_0}{2})}, \ B=A^\dagger\]
where we neglect a real factor common to $A$ and $B$ which will not affect the scattering phase shift.  The phase shift itself has the form:
\begin{equation}
\delta_0 = \frac{1}{2 i} \ln (-\frac{A}{A^\dagger}) 
\label{eq:phaseshift}
\end{equation}
In the limit $ka \ll 1$, the Gamma functions may be expanded in a Taylor series,

\[\Gamma(i k r_0) \simeq \frac{1}{i k r_0} \Gamma(1+ i k r_0) \simeq \frac{1 + i k r_0 \psi(1) } {i k r_0} \Gamma(1) \]
\[\Gamma(\frac{1-2 \lambda}{2} + \frac{i k r_0}{2}) \simeq \Gamma (\frac{1-2 \lambda}{2}) (1+ \frac{i k r_0}{2} \psi^{(0)} (\frac{1-r_0}{2}) )\]
\[\Gamma(1+\frac{2 \lambda}{2} + \frac{i k r_0}{2}) \simeq \Gamma (1+\lambda) (1+ \frac{i k r_0}{2} \psi^{(0)} (1+\lambda)) \]
where $\psi^{(0)}$ is the $m^{th}$-order polygamma function, i.e. $\psi^{(0)}(x) = \frac{d}{dx} \ln \Gamma(x) $.
The final expression for $\delta_0$ now becomes:
\begin{equation}
\delta_0= k r_0 [-\ln(2) + \psi^{(0)}(1) -\frac{1}{2} \psi^{(0)} (\frac{1-2 \lambda}{2}) - \frac{1}{2} \psi^{(0)} (1+ \lambda)]
\label{eq:finaldelta}
\end{equation}
which is related to the s-wave scattering length by the usual formula,
$$\lim_{k\rightarrow 0} k \cot{ \delta_0} = -\frac{1}{a_{sc}}$$
The parameters for the pseudopotential $V_2(r)$ (~\ref{eq:rightpotential}) are hence found by solving the system of equations:

\begin{equation}
a_s = r_0 (-ln(2) + \psi(1) -\frac{1}{2} \psi(\frac{1-2 \lambda}{2}) -\frac{1}{2} \psi(1+\lambda))  
\label{eq:condfirst}
\end{equation}
and
\begin{equation}
EA=-\frac{\hbar^2}{2 \mu {r_0}^2} (\frac{1}{2} \sqrt{\frac{8 \mu V_0 {r_0}^2}{\hbar^2}+1} -\frac{3}{2})^2 
\label{eq:condsecond}
\end{equation}

There are multiple solutions to Eqs. (A8 - A9).  
However, since it is known that Rb${}^{-}$ has only one bound state, 
we impose the additional restriction:
\[N_{\text{bound}} = \lfloor \frac{1}{4} \sqrt{\frac{8 \mu V_0 r_0^2}{\hbar^2} +1} + \frac{1}{4}\rfloor =1\]
The solutions to Eqs. (A8-A9) yield $r_0=9.004786$ a.u. and $V_0=0.061675$ a.u.

\section{Dipole and Non-Adiabatic Coupling Matrix Elements}

Dipole elements take the usual form:
\[\vec{d}_{mn}=\bra{\psi_m} r \cos{\theta} \ket{\psi_n}\]
Non-adiabatic coupling matrix elements are computed via the finite difference formula:
\[A_{mn}=\frac{1}{2 \Delta R} (\gamma_{mn}(R|R+\Delta R) - \gamma_{mn}(R|R-\Delta R))\]
where
\[\gamma_{mn}(R|R\pm \Delta R)=\bra{\psi_m(r|R)}\ket{\psi_n(r|R\pm \Delta R)} \]

\end{document}